\documentclass[a4paper,fleqn,usenatbib]{mnras}

\usepackage{newtxtext,newtxmath}
\usepackage{mathptmx}
\usepackage{txfonts}

\usepackage[T1]{fontenc}
\usepackage{ae,aecompl}

\usepackage{graphicx}	
\usepackage{amsmath}	
\usepackage{amssymb}	

\title[X-ray photopumping in a solar flare]{X-ray line coincidence photopumping in a solar flare}

\author[F. P. Keenan et al.]{
F. P. Keenan,$^{1}$\thanks{E-mail: f.keenan@qub.ac.uk}
K. Poppenhaeger,$^{1}$
M. Mathioudakis,$^{1}$ S. J. Rose,$^{2}$ J. Flowerdew,$^{2}$ 
\newauthor
D. Hynes,$^{2}$ 
D. J. Christian,$^{3}$ J. Nilsen$^{4}$ and W. R. Johnson$^{5}$
\\
$^{1}$Astrophysics Research Centre, School of Mathematics and Physics, Queen's University Belfast, BT7 1NN, UK
\\
$^{2}$Blackett Laboratory, Imperial College, London SW7 2AZ, UK
\\
$^{3}$Department of Physics and Astronomy, California State University, 
Northridge, CA 91330, USA
\\
$^{4}$Lawrence Livermore National Laboratory, P.O. Box 808, Livermore, CA 94551-0808, USA
\\
$^{5}$Department of Physics, University of Notre Dame, Notre Dame, IN 46556, USA
}

\date{Accepted XXX. Received YYY; in original form ZZZ}

\pubyear{2017}

\begin{document}
\label{firstpage}
\pagerange{\pageref{firstpage}--\pageref{lastpage}}
\maketitle

\begin{abstract}
Line coincidence photopumping is a process where the electrons of an atomic or molecular species are radiatively excited through the absorption of line emission from another species at a coincident wavelength. 
There are many instances of line coincidence photopumping in astrophysical sources at optical and ultraviolet wavelengths, with
the most famous example being Bowen fluorescence (pumping of O\,{\sc iii} 303.80\,\AA\ by He\,{\sc ii}), but none to 
our knowledge in X-rays. However, here we report on a scheme 
where a He-like line of Ne\,{\sc ix} at 11.000\,\AA\ 
is photopumped by He-like Na\,{\sc x} at 11.003\,\AA, which predicts significant intensity enhancement in the Ne\,{\sc ix} 82.76\,\AA\
transition under physical conditions found in solar flare plasmas. A comparison of our theoretical models with published X-ray observations of a solar flare obtained during a rocket flight provides evidence for line enhancement, with
the measured degree of enhancement being consistent with that expected from theory, a truly surprising result. Observations of this enhancement
during flares on stars other than the Sun would provide a powerful new diagnostic tool for 
determining the sizes of flare loops in these distant, spatially-unresolved, astronomical sources.
\end{abstract}




\section{Introduction}

Line coincidence photopumping --- the radiative excitation of a transition in an atomic or molecular species due to line emission 
in another species at a coincident wavelength --- was first recognised in astrophysics by  \citet{bowen34}. This process, termed 
Bowen fluorescence, results from the wavelength coincidence of the 2p$^{2}$ $^{3}$P$_{2}$--2p3d $^{3}$P$_{2}$ transition of
O\,{\sc iii} at 303.80\,\AA\ with the He\,{\sc ii} 1s $^{2}$S$_{1/2}$--2p $^{2}$P$_{1/2,3/2}$ resonance lines, with the latter pumping the former, leading to 
O\,{\sc iii} emission features linked to the 2p3d $^{3}$P$_{2}$ level which are much stronger than would be expected in nebular
plasmas. Since the seminal paper by \citeauthor{bowen34}, 
numerous examples of line coincidence photopumping in astrophysical plasmas have been identified. These include, for example:
pumping of O\,{\sc i} 2s$^{2}$2p$^{4}$ $^{3}$P$_{2}$--2s$^{2}$2p$^{3}$3d $^{3}$D$_{1,2,3}$ at 1025.76\,\AA\
by Lyman-$\beta$, which following cascades leads to enhanced emission in the 
O\,{\sc i} 2s$^{2}$2p$^{4}$ $^{3}$P$_{0,1,2}$--2s$^{2}$2p$^{3}$3s $^{3}$S$_{1}$ triplet lines at $\sim$\,1304\,\AA\ \citep{skelton82};
the 2s$^{2}$2p $^{2}$P$_{3/2}$--2s2p$^{2}$ $^{2}$D$_{5/2}$ transition of C\,{\sc ii} at 1335.71\,\AA\
pumping Cl\,{\sc i} 3s$^{2}$3p$^{5}$ $^{2}$P$_{3/2}$--3s$^{2}$3p$^{4}$4s $^{2}$P$_{1/2}$, resulting in an increase in 
intensity in Cl\,{\sc i}
3s$^{2}$3p$^{5}$ $^{2}$P$_{1/2}$--3s$^{2}$3p$^{4}$4s $^{2}$P$_{1/2}$ at 1351.66\,\AA\ \citep{shine83};
2p$^{6}$3s $^{2}$S$_{1/2}$--2p$^{6}$3p $^{2}$P$_{3/2}$ of Mg\,{\sc ii} at 2795.53\,\AA\ pumping
Mn\,{\sc i}  3d$^{5}$4s$^{2}$ a$^{6}$S$_{5/2}$--3d$^{5}$4s4p y$^{6}$P$_{7/2}$,  in turn enhancing
3d$^{5}$4s$^{2}$ a$^{6}$S$_{5/2}$--3d$^{5}$4s4p z$^{8}$P$_{7/2}$ of Mn\,{\sc i} at 5394.67\,\AA\ \citep{doyle01}.
However, the O\,{\sc iii}/He\,{\sc ii} case remains that at the shortest wavelength \citep{hartman13}. 

Although some schemes have been suggested and 
investigated theoretically \citep{sako03}, there 
has been no detection to our knowledge of photopumping at X-ray wavelengths in a high temperature (coronal)
astrophysical source. Coronal emission is usually driven by collisional excitation, and observing a photopumping scheme in X-rays
(which is by definition radiatively excited) would be extremely interesting, especially as this provides
an independent pathway to determine plasma parameters in remote, spatially-unresolved, astronomical sources (see Section 3).

In contrast to the astrophysical case, 
there has been an enormous amount of research on X-ray line coincidence photopumping schemes in laboratory plasmas, due to the possibility of 
developing X-ray lasers 
via this technique.  Although lasing has not yet been achieved, one experimental setup by \citet{porter92} involving the helium-like
ions of sodium (Na\,{\sc x}) and neon (Ne\,{\sc ix}) did demonstrate 
population inversion, with the potential gain modelled by \citet{nilsen91}.  Given the success of this scheme in the
laboratory, we have chosen to model it for astrophysical environments. Our plasma models are discussed in detail in Section 2, with the theoretical results and comparison with solar flare observations given in Section 3. 

\section{Line coincidence photopumping models} 

In Fig. 1 we show the 
the energy level diagram for the Na\,{\sc x}/Ne\,{\sc ix} photopumping scheme of \citet{porter92}. 
The 1s$^{2}$ $^{1}$S$_{0}$--1s2p $^{1}$P$_{1}$ resonance line of Na\,{\sc x}
 at 11.003\,\AA\
pumps the Ne\,{\sc ix} 1s$^{2}$ $^{1}$S$_{0}$--1s4p $^{1}$P$_{1}$ transition at 11.000\,\AA. In laboratory plasmas, there is collisional
redistribution from the 1s4p $^{1}$P$_{1}$ to other 1s4$\ell$ ($\ell$ = s, d, f) levels due to the high electron density, but this does not happen in the lower density astrophysical case. Instead, electrons in 1s4p $^{1}$P$_{1}$ decay to 1s3s $^{1}$S$_{0}$ or 1s3d $^{1}$D$_{2}$, 
producing
emission lines at 224.34\,\AA\ and 233.52\,\AA, respectively. Subsequently, 1s3s $^{1}$S$_{0}$ and 1s3d~$^{1}$D$_{2}$ 
both decay to 1s2p $^{1}$P$_{1}$, producing 82.76\,\AA\ and 81.58\,\AA, respectively.

\begin{figure*}
		\includegraphics[width=14cm]{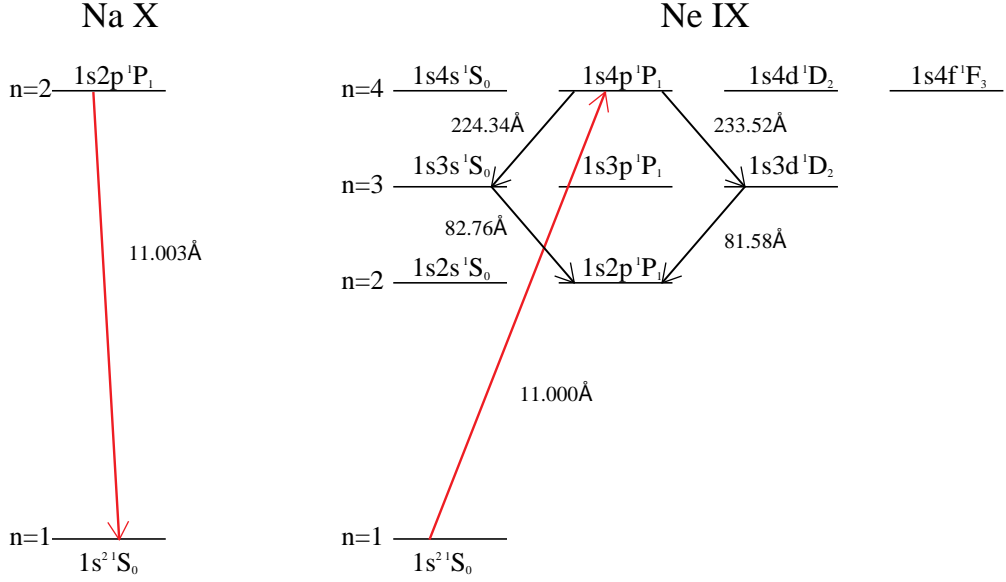}
    \caption{Energy level diagrams for Na\,{\sc x} and Ne\,{\sc ix} showing the X-ray photopumping scheme. The 1s$^{2}$ 
    $^{1}$S$_{0}$--1s4p $^{1}$P$_{1}$ transition of Ne\,{\sc ix} at 11.000\,\AA\ is
photopumped by the 1s$^{2}$ $^{1}$S$_{0}$--1s2p $^{1}$P$_{1}$ resonance line of Na\,{\sc x} at 11.003\,\AA. Subsequently, the 
electrons in 1s4p $^{1}$P$_{1}$ decay to 1s3s $^{1}$S$_{0}$ or 1s3d $^{1}$D$_{2}$, producing the 224.34\,\AA\ and 233.52\,\AA\ emission lines, respectively, 
which in turn both decay to 1s2p $^{1}$P$_{1}$, producing 82.76\,\AA\ and 81.58\,\AA, respectively.
}
\end{figure*}

We  have used the GALAXY modelling code of \citet{rose95} 
to calculate the expected enhancements in the above lines 
of Ne\,{\sc ix} due to photopumping, as a function of
 electron density, N$_{e}$, and plasma pathlength, L.
 Briefly, GALAXY is a time-independent collisional-radiative model that describes the excitation and ionisation within a plasma at a given electron and ion temperature and a given electron number density. It can simulate mixtures of elements through a self-consistent calculation of electron number density. All the atomic data, including energy levels, collisional and radiative excitation and ionisation rates, plus autoionisation and dielectronic recombination rates, are generated internally using simple screened-hydrogenic methods. The description of levels is in the {\em nl}-coupled average-of-configuration approximation. We consider a mixture of H, He, Ne and Na, with the solar coronal abundances of \citet{schmelz12}. 
Using the {\em nl}-coupled configuration average description, GALAXY considers
for Ne and Na the fully stripped ions as well as the H-like configurations 1s, 2s, 2p, 3s, 3p, 3d, \ldots, 4d, 4f; 
the He-like configurations 
1s$^{2}$, 1s2s, 1s2p, 1s3s, 1s3p, 1s3d, \ldots, 1s4d, 1s4f; plus the low-lying Li-like configurations
1s$^{2}$2s, 1s$^{2}$2p. For the elements H and He, the model only includes the fully-stripped ions.
  
GALAXY is a 0-D model; however, two lengths are considered in the calculation of radiation transfer. The first, {\em y}, is the average distance that photons travel to escape, while the second, {\em z}, is the length of plasma along a line-of-sight for which a calculation of the emergent intensity in a spectral line is required. In the calculations reported here, {\em y} is taken as identical to {\em z}.

GALAXY uses a line trapping model to simulate the effect of reabsorption of line radiation. In this case the radiative rate (A-value) is reduced to account for the reabsorption as described by \citet{rose95}, with the difference that the escape factor now used in 
GALAXY differs from that reported in \citeauthor{rose95} by adopting the more accurate description described by \citet{phillips08}.  Line trapping is included on all the electric-dipole allowed transitions from the ground states of H-like and He-like ions. 

The line coincidence photopumping is calculated in GALAXY by including radiative excitation and de-excitation rates between the 
upper level $\beta$ and lower level $\alpha$ (R$^{r}_{\alpha \rightarrow \beta}$ and R$^{r}_{\beta \rightarrow \alpha}$, respectively) by

\vspace*{0.2in}

R$^{r}_{\alpha \rightarrow \beta}$ = A$_{\beta \rightarrow \alpha}${\em n}$_{ph}$(${\Omega_\beta}$/${\Omega_\alpha}$)

\vspace*{0.2in}

R$^{r}_{\beta \rightarrow \alpha}$ = A$_{\beta \rightarrow \alpha}$(1 + {\em n}$_{ph}$)

\vspace*{0.2in}

\noindent
where {\em n}$_{ph}$ is the radiation modal photon density covering the transition $\alpha \longleftrightarrow \beta$, and 
$\Omega_{\alpha}$ and $\Omega_\beta$ are the degeneracies of levels $\alpha$ and $\beta$, respectively. For the case of interest here involving an internal source of photons where 
a transition $\chi \longleftrightarrow \delta$  is coincident with and pumps the transition $\alpha \longleftrightarrow \beta$  
then we make the approximation

\vspace*{0.2in}

{\em n}$_{ph}$ =  1/([{\em n}$_{\chi}$$\Omega_\delta$/{\em n}$_\delta$$\Omega_\chi$] -- 1)

\vspace*{0.2in}

\noindent
where {\em n}$_\chi$ and $\Omega_\chi$ are the ion number density and degeneneracy of level $\chi$, respectively.
This method has been used in previous calculations of line coincidence photopumping, for example by \citet{judge88}
for the pumping of S\,{\sc i}
lines by H\,{\sc i} Ly-$\alpha$ radiation in the chromospheres of giant stars. The use of the above equations 
is approximate, and requires that the $\chi \longleftrightarrow \delta$ line (in our case the Na\,{\sc x} 
1s$^{2}$ $^{1}$S$_{0}$--1s2p $^{1}$P$_{1}$ transition) is optically thick and that this opacity-broadened line effectively overlaps the
$\alpha \longleftrightarrow \beta$  line (in our case Ne\,{\sc ix} 1s$^{2}$ $^{1}$S$_{0}$--1s4p $^{1}$P$_{1}$). In the calculations reported here, Na\,{\sc x} 1s$^{2}$ $^{1}$S$_{0}$--1s2p $^{1}$P$_{1}$ is optically thick for cases showing a significant line enhancement ratio, and the line centres of the 
Na\,{\sc x} 1s$^{2}$ $^{1}$S$_{0}$--1s2p $^{1}$P$_{1}$ and Ne\,{\sc ix} 1s$^{2}$ $^{1}$S$_{0}$--1s4p $^{1}$P$_{1}$ transitions are close to the Doppler width of each line at the electron temperatures considered. Consequently, we consider that our modelling of the line coincidence photopumping is adequate in approximately identifying the effect. 

GALAXY calculates the emission intensity integrated over a spectral line along the line-of-sight in the plasma of length {\em z} using the expressions given in  \citet{rose95}. We believe that the approximations made in terms of the level of detail in the description of the atomic physics and of the radiation transfer are adequate to provide an indication of the effect of line coincidence photopumping in this situation. However, more accurate calculations would be needed to allow any observed line enhancements to be used reliably to predict plasma properties, which we will report in future publications.

\section{Results and discussion}

We have calculated the line enhancement factors (i.e. line intensity of the photopumped line divided by that 
when there is no pumping) for several Ne\,{\sc ix} transitions for a range of 
N$_{e}$ (= 10$^{10}$--10$^{13}$\,cm$^{-3}$) and L (= 10$^{9}$--10$^{13}$\,cm) appropriate to flaring coronal loops
 in the Sun \citep{shibata11} and other active, late-type stars \citep{mullan06}. Our results for the 82.76\,\AA\ line are shown
 in Fig. 2. Predicted enhancements
 in the 224.34\,\AA\ and 233.52\,\AA\ lines are similar to those for 82.76\,\AA, while for 81.58\,\AA\ they are about a factor of 
 20 smaller. An inspection of Fig. 2 reveals that the enhancement factor 
 does not depend strongly on the electron density unless N$_{e}$ $\geq$ 10$^{11}$\,cm$^{-3}$, but is 
 sensitive to the pathlength for 
 L $\geq$ 10$^{10}$\,cm. 
We note that the line enhancement factors are not very sensitive to the 
adopted electron temperature, T$_{e}$. For example, for a pathlength L = 10$^{11}$\,cm, increasing the temperature from that of maximum
Ne\,{\sc ix} fractional abundance in ionisation equilibrium, T$_{e}$ = 1.6$\times$10$^{6}$\,K \citep{bryans09}, to 2.3$\times$10$^{6}$\,K, leads to
a change in the line enhancement factor of less than 20 per cent at an electron density of N$_{e}$ = 10$^{11}$\,cm$^{-3}$, 
decreasing to less than 15 per cent at N$_{e}$ = 10$^{13}$\,cm$^{-3}$.

\begin{figure*}
		\includegraphics[width=14cm]{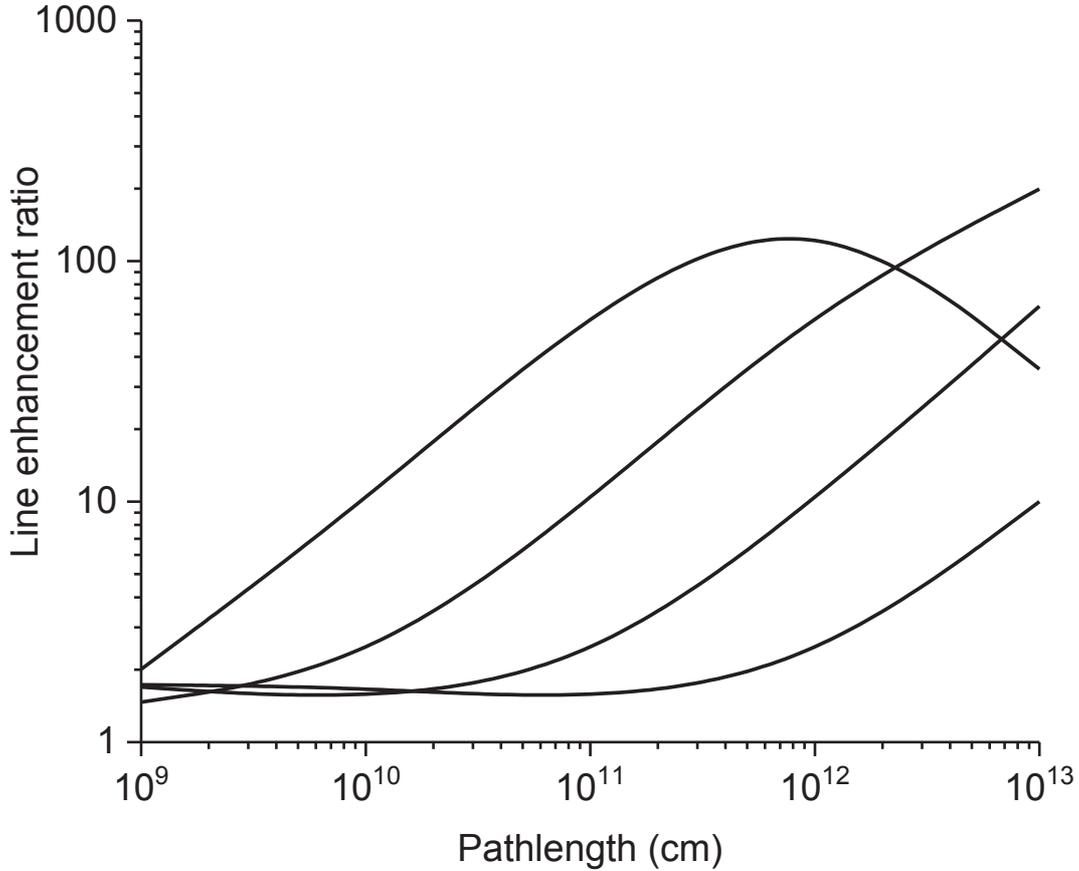}
    \caption{Plot of the enhancement factor (i.e. line intensity of the photopumped line divided by that when there is no pumping) for the 82.76\,\AA\ line of Ne\,{\sc ix}, calculated with the GALAXY code \citep{rose95} as a function of pathlength L for values of electron density of (from 
bottom to top) N$_{e}$ = 10$^{10}$, 10$^{11}$, 10$^{12}$ and 10$^{13}$\,cm$^{-3}$. The adopted electron temperature is that of 
maximum Ne\,{\sc ix} fractional abundance in ionisation equilibrium, T$_{e}$ = 1.6$\times$10$^{6}$\,K \citep{bryans09}, 
although we note that the results are not very sensitive to T$_{e}$ (see Section 3). 
The decrease in enhancement at large L for the N$_{e}$ = 10$^{13}$\,cm$^{-3}$ curve is due to the onset of opacity in the 82.76\,\AA\ line.
}
\end{figure*}

 In solar and late-type stellar flare spectra, the Ne\,{\sc ix} 224.34\,\AA\ and 233.52\,\AA\ lines are unfortunately blended with the strong Fe\,{\sc xiv}
 224.35\,\AA\ and O\,{\sc iv} 233.55\,\AA\ features, respectively. A synthetic solar flare spectrum calculated with the latest version (8.0.1) of the CHIANTI database \citep{dere97, delzanna15} with the coronal abundances of \citet{schmelz12}
 indicates that  Ne\,{\sc ix} makes a contribution of less than 1 per cent to the total line flux in both instances, while solar flare observations confirm that the features are due to Fe\,{\sc xiv} \citep{bhatia94} and O\,{\sc iv} \citep{widing82}. In the case of the 
 81.58\,\AA\ transition, the predicted enhancement is too small to detect the feature, even under optimal conditions.

However, the situation is different for the Ne\,{\sc ix} 1s2p $^{1}$P$_{1}$--1s3s $^{1}$S$_{0}$ transition at 82.76\,\AA. Although this lies in a relatively unexplored spectral region, one very interesting dataset is a soft X-ray rocket spectrum of a solar flare between 10 -- 95\,\AA\ obtained 
at a high spectral resolution of 0.02\,\AA\ by \citet{acton85}. In particular, \citeauthor{acton85} note an unidentified emission line at 82.76\,\AA, which is listed as blended.
Assuming an uncertainty of $\pm$\,5 in the digit beyond the last significant figure in the National Institute of Standards and Technology (NIST) Atomic Spectra Database \citep{kramida16},  
one obtains a value 82.760\,$\pm$\,0.001\,\AA\ for the wavelength of the 1s2p $^{1}$P$_{1}$--1s3s $^{1}$S$_{0}$ transition of Ne\,{\sc ix}. 
However, we have undertaken new relativistic many-body calculations, including the Breit interaction and radiative corrections \citep{chen93, chen01, cheng94}
but with more basis functions, which
lead to a value of 82.762\,$\pm$\,0.001\,\AA. 
Averaging these two results gives a final predicted wavelength of 82.761\,$\pm$\,0.001\,\AA, confirming the NIST listing.

The CHIANTI synthetic flare spectrum \citep{dere97, delzanna15} indicates that, even in the absence of photopumping, 
the Ne\,{\sc ix} 82.76\,\AA\ transition should dominate the emission feature observed by \citet{acton85}, 
contributing at least 50 per cent to the total line flux, with Fe\,{\sc xx} and Fe\,{\sc xxii} providing 
the main blending transitions.
Furthermore, the \citeauthor{acton85} flare spectrum also contains the Ne\,{\sc ix} 1s$^{2}$ $^{1}$S$_{0}$--1s2p $^{1}$P$_{1}$ resonance line at
13.45\,\AA\ and 
1s$^{2}$ $^{1}$S$_{0}$--1s2s $^{3}$S$_{1}$ forbidden line at 13.70\,\AA. The electron density of the flare is 
N$_{e}$ $\sim$ 10$^{11}$\,cm$^{-3}$ \citep{brown86}, 
for which CHIANTI predicts line intensity ratios (in photon units) of 82.76/13.45 = 0.019 and 82.76/13.70 = 0.020, compared to experimental values of 0.20 and 0.24, respectively. We note that the measured 
13.45/13.70 line intensity ratio and the theoretical estimate from CHIANTI  are both 1.2, indicating no blending in these features. 
Hence, assuming that Ne\,{\sc ix} is responsible for $\sim$\,50 per cent of the 82.76\,\AA\ line intensity, both the 82.76/13.45 and
82.76/13.70  ratios
indicate that 82.76\,\AA\ is enhanced by a factor of $\sim$\,5. The maximum flare loop length for the Sun is observed to be
L $\sim$ 10$^{10}$\,cm \citep{shibata11}, which from Fig. 2 at N$_{e}$ = 10$^{11}$\,cm$^{-3}$  indicates an expected enhancement of a factor of $\sim$\,2, not too different from the measured value.

Although the above solar flare results are very encouraging, providing to our knowledge the first evidence for X-ray
line coincidence photopumping in an astrophysical source, they must be treated with some caution. 
The flare spectrum of \citet{acton85}, recorded on photographic film, is no longer accessible and hence the quality of the 82.76\,\AA\ line measurement (and indeed those of other transitions) cannot be confirmed. In addition, the 13.45 + 13.70\,\AA\ and 82.76\,\AA\ features lie close to opposite ends of the flare wavelength coverage (10 -- 95\,\AA), so that instrumental sensitivity calibration may be an issue. 

Clearly, further observations of the Ne\,{\sc ix} 13.45, 13.70 and 82.76\,\AA\ lines are desirable to unreservedly confirm our findings. Ideally,
this would be 
for flare 
plasmas with larger values of electron density and pathlength than the solar case, as Fig. 2 indicates these would show even greater enhancement factors. Such plasmas are provided by, for example, late-type stellar coronal sources, many of which are believed to 
have high density (N$_{e}$ $\geq$ 10$^{11}$\,cm$^{-3}$), large pathlength (L $\sim$\,10$^{10}$--10$^{12}$\,cm) flaring loops
\citep{mullan06}.
Detection of 82.76\,\AA\ line enhancement in these sources, as well as being an
extremely interesting plasma effect, would also have a profound astrophysical application.  The electron density of a flare plasma is relatively straightforward to determine from spectra in the $\sim$\,13--83\,\AA\ range covering the Ne\,{\sc ix} features,
due to the presence of numerous N$_{e}$-diagnostic emission lines \citep{brown86}. 
By contrast, the methods developed for determining 
stellar flare loop pathlengths L require various assumptions to be made. For example,
that of \citet{haisch83} 
assumes no additional flare heating during flare decay, while the method of \citet{shibata02} 
requires an assumption of the value for the magnetic field strength (see \citealt{mullan06} for 
more details on these and other methods). A comparison of the 82.76\,\AA\ line enhancement factor and flare electron density 
measured from an X-ray spectrum with the results in Fig. 2 will give an independent determination of L, and in particular will yield values for high L stellar coronal plasmas, as the enhancement factor is predicted to become very sensitive to L under such conditions.  
These measurements of L may be compared with those derived using other methods, to determine the reliability of the latter and the assumptions made. 
Indeed, in principle the enhancement effect could provide a very powerful diagnostic tool for determining plasma pathlengths 
in any remote, spatially-unresolved, astrophysical source which shows similar values of N$_{e}$ and L to the late-type stellar cases, such as the coronal regions of active galactic nuclei \citep{reeves16}.

Unfortunately, the full wavelength range covering the Ne\,{\sc ix} 13.45, 13.70 and 82.76\,\AA\ lines is not accessible with most operating telescopes. The
only currently operational instrument covering this range is the Chandra
LETG spectrograph in conjunction with the HRC-S detector \citep{brinkman00}. However, while
its spectral resolution is sufficiently high to separate the Ne\,{\sc ix} lines at
13.45 and 13.70\,\AA, it cannot resolve blends of these with
surrounding Fe\,{\sc xix} lines in flare observations \citep{ness03}. 
Future X-ray observatories may provide better
opportunities to study the Ne\,{\sc ix} photopumping scheme in astrophysical
sources; a spectral resolution of better than 0.02\,\AA\ and a high throughput will
be required to determine the behaviour of the Ne\,{\sc ix} 13.45, 13.70
and 82.76\,\AA\ features in a time-resolved manner during stellar flares. While the
planned capabilities for Athena+ include high resolution spectral
capabilities only at wavelengths below Ne\,{\sc ix} 82.76\,\AA\ \citep{nandra13}, other
missions which are yet to be proposed, such as LynX/X-ray Surveyor \citep{gaskin15}, may be able to
provide the necessary observational features.

\section*{Acknowledgements}

FPK, KP and MM are grateful to the Science and Technology Facilities Council for financial support.
The research leading to these results has received funding from the European Community's Seventh Framework Programme (FP7/2007--2013) under grant agreement no. 606862 (F-CHROMA).
The work of JN was performed under the auspices of the US Department of Energy by Lawrence Livermore National Laboratory under Contract DE-AC52-07NA27344. We thank K. T. Cheng for providing valuable advice concerning the atomic physics calculations.
CHIANTI is a collaborative project involving George Mason University, the University of Michigan (USA) and the University of Cambridge (UK). 









\bsp	
\label{lastpage}
\end{document}